\documentstyle[12pt,epsf,axodraw]{article}

\newcommand{\beq}[1]{
\begin{equation}\label{#1}}
\newcommand{\eeq}{\end{equation}}
\newcommand{\bea}[1]{
\begin{eqnarray}\label{#1}}
\newcommand{\eea}{\end{eqnarray}}
\begin{document}

\title{\Large \bf Electroproduction of two light vector mesons in next-to-leading
BFKL\footnote{The content of this contribution is based on Ref.~\cite{IP05}, 
to which we refer for additional details and for a more exhaustive list of 
references.}}
\author{D.Yu. Ivanov$^1$ and A. Papa$^2$ \bigskip \\
{\it $^1$~Sobolev Institute of Mathematics, 630090 Novosibirsk, Russia} \\  
{\it $^2$~Dipartimento di Fisica, Universit\`a
della Calabria} \\
{\it and Istituto Nazionale di Fisica Nucleare, Gruppo collegato di Cosenza} \\
{\it I-87036 Arcavacata di Rende, Cosenza, Italy}}

\maketitle


{\bf Abstract}

\noindent We calculate the amplitude for the forward electroproduction of two light
vector mesons in next-to-leading order BFKL. This amplitude
represents the first next-to-leading order amplitude ever calculated for a
collision process between strongly interacting colorless particles.

\section{Introduction}

It is believed that the ``gold-plated'' measurement for the possible realization
of the BFKL dynamics~\cite{BFKL} is the $\gamma^* \gamma^*$ total cross section. 
However, a prediction for this cross section is not yet available with 
next-to-leading accuracy since the calculation of one necessary ingredient, the 
$\gamma^*$ to $\gamma^*$ impact factor in the next-to-leading order, has not yet 
been completed after year-long efforts~\cite{gammaIF}.
Here we propose as ``silver-plated'' measurement the differential cross section
for the $\gamma^* \gamma^*$ to two light vector mesons process in the 
forward case, i.e. for minimum momentum transfer.

\begin{figure}[tb]
\centering
\setlength{\unitlength}{0.35mm}
\begin{picture}(300,200)(0,0)

\Photon(0,190)(100,190){3}{7}
\ArrowLine(200,190)(300,190)
\Text(50,200)[c]{$p$}
\Text(250,200)[c]{$p_1$}
\Text(150,190)[]{$\Phi_1(\vec q_1, s_0)$}
\Oval(150,190)(20,50)(0)

\ZigZag(125,174)(125,120){5}{7}
\ZigZag(175,174)(175,120){5}{7}
\ZigZag(125,26)(125,80){5}{7}
\ZigZag(175,26)(175,80){5}{7}

\ArrowLine(110,160)(110,130)
\ArrowLine(190,130)(190,160)
\ArrowLine(110,70)(110,40)
\ArrowLine(190,40)(190,70)

\Text(100,145)[r]{$q_1$}
\Text(200,145)[l]{$q_1$}
\Text(100,55)[r]{$q_2$}
\Text(200,55)[l]{$q_2$}

\GCirc(150,100){30}{1}
\Text(150,100)[]{$G(\vec q_1,\vec q_2)$}

\Photon(0,10)(100,10){3}{7}
\ArrowLine(200,10)(300,10)
\Text(50,0)[c]{$p'$}
\Text(250,0)[c]{$p_2$}
\Text(150,10)[]{$\Phi_2(-\vec q_2,s_0)$}
\Oval(150,10)(20,50)(0)

\end{picture}

\caption[]{Schematic representation of the amplitude for the $\gamma^*(p)\,
\gamma^*(p') \to V(p_1)\, V(p_2)$ forward scattering.}
\label{fig:BFKL}
\end{figure}
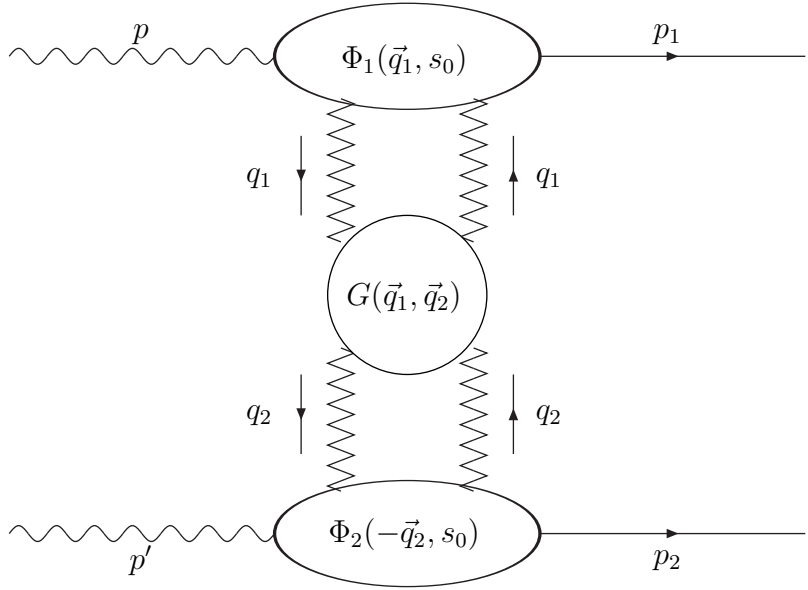

In the BFKL approach, both in the leading logarithmic approximation (LLA),
which means resummation of leading energy logarithms,
all terms $(\alpha_s\ln(s))^n$, and in the
next-to-leading approximation (NLA), which means resummation of all terms
$\alpha_s(\alpha_s\ln(s))^n$, the (imaginary part of the) amplitude for
a large-$s$ hard collision process can be written as the convolution of the
Green's function of two interacting Reggeized gluons with the impact factors
of the colliding particles (see, for example, Fig.~\ref{fig:BFKL}).

The NLA singlet BFKL Green's function for the forward case of interest here
is known since several years~\cite{NLA-kernel}. Moreover, the impact factor 
for the transition from a virtual photon $\gamma^*$ to a light neutral vector meson 
$V=\rho^0, \omega, \phi$ has been recently calculated in the NLA in the
forward case, up to contributions suppressed as inverse powers of the photon 
virtuality~\cite{IKP04}. Therefore we have all is needed to build
the NLA amplitude for the $\gamma^* \gamma^* \to V V$ reaction. 

The knowledge of this amplitude is interesting first of all for theoretical 
reasons, since it could shed light on the role and the optimal choice of the 
energy scales entering the BFKL approach.
Moreover, it could be used as a test-ground for comparisons with approaches
different from BFKL, such as DGLAP, and with possible next-to-leading order
extensions of other approaches, such as color dipole and $k_t$-factorization.
But it could be interesting also for the possible applications to the
phenomenology. Indeed, the calculation of the $\gamma^* \to V$ impact factor is
the first step towards the application of BFKL approach to the description of
processes such as the vector meson electroproduction $\gamma^* p\to V p$, being
carried out at the HERA collider, and the production of two mesons in the photon
collision, $\gamma^*\gamma^*\to VV$ or $\gamma^* \gamma \to VJ/\Psi$, which can be
studied at high-energy $e^+e^-$ and $e\gamma$ colliders.

The process considered here has been studied recently in the Born (2-gluon exchange)
limit in~\cite{PSW} and in the LLA~\cite{PSW1} for arbitrary transverse momentum. 
In Ref.~\cite{PSW1} also an estimate of NLA effects has been given in the forward 
case.

\section{The NLA amplitude}

We consider the production of two light vector mesons ($V=\rho^0, \omega, \phi$) in
the collision of two virtual photons,
\beq{process}
\gamma^*(p) \: \gamma^*(p')\to V(p_1) \:V(p_2) \;.
\eeq
Here, $p_1$ and $p_2$ are taken as Sudakov vectors satisfying $p_1^2=p_2^2=0$ and
$2(p_1 p_2)=s$; the virtual photon momenta are instead
\beq{kinphoton}
p =\alpha p_1-\frac{Q_1^2}{\alpha s} p_2 \;, \hspace{2cm}
p'=\alpha^\prime p_2-\frac{Q_2^2}{\alpha^\prime s} p_1 \;,
\eeq
so that the photon virtualities turn to be $p^2=-Q_1^2$ and $(p')^2=-Q_2^2$.
We consider the kinematics when
\beq{kin}
s\gg Q^2_{1,2}\gg \Lambda^2_{QCD} \, ,
\eeq
and
\beq{alphas}
\alpha=1+\frac{Q_2^2}{s}+{\cal O}(s^{-2})\, , \quad
\alpha^\prime =1+\frac{Q_1^2}{s}+{\cal O}(s^{-2})\, .
\eeq 
In this case vector mesons are produced by longitudinally polarized photons in
the longitudinally polarized state~\cite{IKP04}. Other helicity amplitudes are
power suppressed, with a suppression factor $\sim m_V/Q_{1,2}$.
We will discuss here the amplitude of the forward scattering, i.e.
when the transverse momenta of produced $V$ mesons are zero or 
when the variable $t=(p_1-p)^2$ takes its maximal value $t_0=-Q_1^2Q_2^2/s+{\cal
O}(s^{-2})$.

The forward amplitude in the BFKL approach may be presented as follows
\beq{imA}
{\cal I}m_s\left( {\cal A} \right)=\frac{s}{(2\pi)^2}\int\frac{d^2\vec q_1}{\vec
q_1^{\,\, 2}}\Phi_1(\vec q_1,s_0)\int
\frac{d^2\vec q_2}{\vec q_2^{\,\,2}} \Phi_2(-\vec q_2,s_0)
\int\limits^{\delta +i\infty}_{\delta
-i\infty}\frac{d\omega}{2\pi i}\left(\frac{s}{s_0}\right)^\omega
G_\omega (\vec q_1, \vec q_2)\, .
\eeq
This representation for the amplitude is valid with NLA accuracy.
The scale $s_0$ is artificial. It is introduced in the BFKL approach 
to perform the Mellin transform from the $s$-space to the complex angular
momentum plane and must disappear in the full expression for the amplitude
at each fixed order of approximation. Using the result for the meson
NLA impact factor such cancellation was demonstrated explicitly in
Ref.~\cite{IKP04} for the process in question.

In Eq.~(\ref{imA}), $\Phi_{1}(\vec q_1,s_0)$ and $\Phi_{2}(-\vec q_2,s_0)$
are the impact factors describing the transitions $\gamma^*(p)\to V(p_1)$
and $\gamma^*(p')\to V(p_2)$, respectively. They are presented as an expansion in 
$\alpha_s$
\beq{impE}
\Phi_{1,2}(\vec q)= \alpha_s \,
D_{1,2}\left[C^{(0)}_{1,2}(\vec q^{\,\, 2})+\bar\alpha_s
C^{(1)}_{1,2}(\vec
q^{\,\, 2})\right] \, , \quad D_{1,2}=-\frac{4\pi e_q  f_V}{N_c Q_{1,2}}
\sqrt{N_c^2-1}\, ,
\eeq
where ${\bar \alpha_s}=\alpha_s N_c/\pi$, $f_V$ is the meson dimensional 
coupling constant ($f_{\rho}\approx 200\, \rm{ MeV}$) and $e_q$ should be 
replaced by $e/\sqrt{2}$, $e/(3\sqrt{2})$ and $-e/3$ for the case of $\rho^0$, 
$\omega$ and $\phi$ meson production, respectively.
In the collinear factorization approach the meson transition impact factor
is given as a convolution of the hard scattering amplitude for the
production of a collinear quark--antiquark pair with the meson distribution
amplitude (DA). The integration variable in this convolution is the fraction $z$
of the meson momentum carried by the quark ($\bar z\equiv 1-z$ is
the momentum fraction carried by the antiquark):
\beq{imps1}
C^{(0)}_{1,2}(\vec q^{\,\, 2})=\int\limits^1_0 dz \,
\frac{\vec q^{\,\, 2}}{\vec q^{\,\, 2}+z \bar zQ_{1,2}^2}\phi_\parallel (z)
\, .
\eeq
The NLA correction to the hard scattering amplitude, for a photon with virtuality
equal to $Q^2$, is defined as follows
\beq{imps2}
C^{(1)}(\vec q^{\,\, 2})=\frac{1}{4 N_c}\int\limits^1_0 dz \,
\frac{\vec q^{\,\, 2}}{\vec q^{\,\, 2}+z \bar zQ^2}[\tau(z)+\tau(1-z)]
\phi_\parallel (z)
\, ,
\eeq
with $\tau(z)$ given in the Eq.~(75) of Ref.~\cite{IKP04}.
$C^{(1)}_{1,2}(\vec q^{\,\, 2})$ are given by the previous expression with
$Q^2$ replaced everywhere in the integrand by $Q^2_1$ and $Q^2_2$,
respectively.
Below we will use the DA in the asymptotic form, $\phi_\parallel^{as}=6z(1-z)$,
both for the simplicity of the presentation and because, according to QCD sum 
rules estimates~\cite{BRAUN}, $\phi_\parallel^{as}$ may be indeed a 
good approximation for the DA of light vector mesons. 

The Green's function in (\ref{imA}) is determined by the BFKL equation
\beq{Green}
\delta^2(\vec q_1-\vec q_2)=\omega \, G_\omega (\vec q_1, \vec q_2)-
\int d^2 \vec q \, K(\vec q_1,\vec q)\, G_\omega (\vec q, \vec q_2) \;,
\eeq
where $K(\vec q_1,\vec q_2)$ is the BFKL kernel. In the transverse momentum 
representation (see~\cite{IP05} for details), it can be written as
\beq{kern}
\hat K=\bar \alpha_s \hat K^0 + \bar \alpha_s^2 \hat K^1\;,
\eeq
where $\hat K^0$ is the BFKL kernel in the LLA, $\hat K^1$ represents the NLA 
correction.

The basis of eigenfunctions of the LLA kernel,
\beq{KLLA}
\hat K^0 |\nu\rangle = \chi(\nu)|\nu\rangle \, , \;\;\;\;\;\;\;\;\;\;
\chi (\nu)=
2\psi(1)-\psi\left(\frac{1}{2}+i\nu\right)-\psi\left(\frac{1}{2}-i\nu\right)\, ,
\eeq
is given by the following set of functions:
\beq{nuLLA}
\langle\vec q\, |\nu\rangle =\frac{1}{\pi \sqrt{2}}\left(\vec q^{\,\, 2}\right)
^{i\nu-\frac{1}{2}} \;, \hspace{2cm}
\langle \nu^\prime | \nu\rangle =\int \frac{d^2\vec q}
{2 \pi^2 }\left(\vec q^{\,\, 2}\right)
^{i\nu-i\nu^\prime -1}=\delta(\nu-\nu^\prime)\, .
\eeq
The action of the full NLA BFKL kernel on these functions may be expressed
as follows:
\bea{Konnu}
\hat K|\nu\rangle &=&
\bar \alpha_s(\mu_R) \chi(\nu)|\nu\rangle
 +\bar \alpha_s^2(\mu_R)
\left(\chi^{(1)}(\nu)
+\frac{\beta_0}{4N_c}\chi(\nu)\ln(\mu^2_R)\right)|\nu\rangle
\nonumber \\
&+& \bar
\alpha_s^2(\mu_R)\frac{\beta_0}{4N_c}\chi(\nu)\left(i\frac{\partial}{\partial \nu}
\right)|\nu\rangle \;,
\eea
where the first term represents the action of LLA kernel, while the second
and the third ones stand for the diagonal and the non-diagonal parts of the
NLA kernel. We refer to Ref.~\cite{IP05} for the expression of $\bar\chi(\nu)$.

We will need also the $|\nu\rangle$ representation for the impact factors, which is
defined by the following expressions
\beq{nuu}
\frac{C_1^{(0)}(\vec q^{\,\, 2})}{\vec
q^{\,\, 2}}=\int\limits_{-\infty}^{+\infty}\, d\, \nu^\prime \,c_1(\nu^\prime)
\langle\nu^\prime| \vec q\rangle \;, \hspace{2cm}
\frac{C_2^{(0)}(\vec q^{\,\, 2})}{\vec q^{\,\, 2}}=\int\limits_{-\infty}^{+\infty}
\, d\, \nu \,c_2(\nu)\,\langle\vec q|\nu\rangle \;,
\eeq
\beq{imp1}
c_1(\nu)=\int d^2\vec q \,\, C_1^{(0)}(\vec q^{\, 2})
\frac{\left(\vec q^{\, 2}\right)^{i\nu-\frac{3}{2}}}{\pi \sqrt{2}}
\, ,\;\;\;\;\;
c_2(\nu)=\int d^2\vec q \,\, C_2^{(0)}(\vec q^{\, 2})
\frac{\left(\vec q^{\, 2}\right)^{-i\nu-\frac{3}{2}}}{\pi \sqrt{2}} \, ,
\eeq
and by similar equations for $c_1^{(1)}(\nu)$ and $c_2^{(1)}(\nu)$
from the NLA corrections to the impact factors, $C_1^{(1)}(\vec
q^{\,\, 2})$ and $C_2^{(1)}(\vec q^{\,\, 2})$.

Using the above formulas one can derive, after some algebra,
the following representation for the amplitude
\[
\frac{{\cal I}m_s\left( {\cal A} \right)}{D_1D_2}=\frac{s}{(2\pi)^2}
\int\limits^{+\infty}_{-\infty}
d\nu \left(\frac{s}{s_0}\right)^{\bar \alpha_s(\mu_R) \chi(\nu)}
\alpha_s^2(\mu_R) c_1(\nu)c_2(\nu)\left[1+\bar \alpha_s(\mu_R)
\left(\frac{c^{(1)}_1(\nu)}{c_1(\nu)}
+\frac{c^{(1)}_2(\nu)}{c_2(\nu)}\right)
\right.
\]
\beq{amplnla}
\left.
+\bar \alpha_s^2(\mu_R)\ln\left(\frac{s}{s_0}\right)
\left(\bar
\chi(\nu)+\frac{\beta_0}{8N_c}\chi(\nu)\left[-\chi(\nu)+\frac{10}{3}
+i\frac{d\ln(\frac{c_1(\nu)}{c_2(\nu)})}{d\nu}+2\ln(\mu_R^2)\right]
\right)\right] \; .
\eeq
We find that
\beq{impsnu}
c_{1,2}(\nu)=
\frac{\left(Q^2_{1,2}\right)^{\pm i\nu-\frac{1}{2}}}{\sqrt{2}}
\frac{\Gamma^2 [\frac{3}{2}\pm i\nu]}{\Gamma [3\pm 2i\nu]}
\frac{6\pi}{\cosh (\pi
\nu)}\, .
\eeq

Using Eq.~(\ref{amplnla}) we construct the series representation for the
amplitude
\bea{series}
\frac{Q_1Q_2}{D_1 D_2} \frac{{\cal I}m_s {\cal A}}{s} &=&
\frac{1}{(2\pi)^2}  \alpha_s(\mu_R)^2 \label{honest_NLA} \\
& \times &
\biggl[ b_0
+\sum_{n=1}^{\infty}\bar \alpha_s(\mu_R)^n   \, b_n \,
\biggl(\ln\left(\frac{s}{s_0}\right)^n   +
d_n(s_0,\mu_R)\ln\left(\frac{s}{s_0}\right)^{n-1}     \biggr)
\biggr]\;, \nonumber
\eea
where the coefficients
\beq{bs}
\frac{b_n}{Q_1Q_2}=\int\limits^{+\infty}_{-\infty}d\nu \,  c_1(\nu)c_2(\nu)
\frac{\chi^n(\nu)}{n!} \, ,
\eeq
are determined by the kernel and the impact factors in LLA.
The expression for the coefficients $d_n$ can be easily determined from 
Eq.~(\ref{amplnla}) and is given in Ref.~\cite{IP05}.

One should stress that both representations of the amplitude~(\ref{series})
and~(\ref{amplnla}) are equivalent with NLA accuracy, since they differ only by
next-to-NLA (NNLA) terms. 

It can be easily shown that the amplitude~(\ref{series}) is 
independent in the NLA from the choice of energy and strong coupling
scales. It is also possible to trace the contributions to each $d_n$ coefficient 
coming from the NLA corrections to the BFKL kernel and from the NLA impact factors
(see Ref.~\cite{IP05} for details).

\section{Numerical results}

In this Section we present some numerical results for the amplitude given
in Eq.~(\ref{series}) for the $Q_1=Q_2\equiv Q$ kinematics, i.e. in the
``pure'' BFKL regime. The other interesting regime, $Q_1\gg Q_2$ or vice-versa,
where collinear effects could come heavily into the game, will not be 
considered here. We will emphasize in particular the dependence on 
the renormalization scale $\mu_R$ and $s_0$ in the NLA result.

In all the forthcoming figures the quantity on the vertical axis is the
L.H.S. of Eq.~(\ref{series}), ${\cal I}m_s ({\cal A})Q^2/(s \, D_1 D_2)$.
In the numerical analysis presented below we truncate the series in the
R.H.S. of Eq.~(\ref{series}) to $n=20$, after having verified that
this procedure gives a very good approximation of the infinite sum
for the $Y$ values $Y\leq 10$. We use the two--loop running coupling
corresponding to the value $\alpha_s(M_Z)=0.12$.

We have calculated numerically the $b_n$ and $d_n$ coefficients for
$n_f=5$ and $s_0=Q^2=\mu_R^2$, getting
\beq{coe}
\begin{array}{lllll}
b_0=17.0664  & b_1=34.5920   & b_2=40.7609  & b_3=33.0618   &
b_4=20.7467  \\
             & b_5=10.5698  & b_6=4.54792 & b_7=1.69128  &
b_8=0.554475\\
& & & & \\
& d_1=-3.71087 & d_2=-11.3057 & d_3=-23.3879 & d_4=-39.1123 \\
& d_5=-59.207 & d_6=-83.0365 & d_7=-111.151 & d_8=-143.06 \;. \\
\end{array}
\eeq
In this case contributions to the $d_n$ coefficients originating from
the NLA corrections to the impact factors are
\beq{coeffim}
\begin{array}{lllll}
& d_1^{\rm{imp}}=-3.71087 & d_2^{\rm{imp}}=-8.4361 & d_3^{\rm{imp}}=-13.1984 &
d_4^{\rm{imp}}=-18.0971 \\
& d_5^{\rm{imp}}=-23.0235 & d_6^{\rm{imp}}=-27.9877 &
d_7^{\rm{imp}}=-32.9676 &
d_8^{\rm{imp}}=-37.9618 \;. \\
\end{array}
\eeq
Thus, comparing~(\ref{coe}) and (\ref{coeffim}), we see that the contribution
from the kernel starts to be larger than the impact factor one only for $n\geq 4$.

These numbers make visible the effect of the NLA corrections: the $d_n$
coefficients are negative and increasingly large in absolute values as the
perturbative order increases. The NLA corrections turn to be very large.
In this situation the optimization of perturbative expansion, in our case the
choice of the renormalization scale $\mu_R$ and of the energy scale $s_0$,
becomes an important issue.
Below we will adopt the principle of minimal sensitivity (PMS)~\cite{Stevenson}.
Usually PMS is used to fix the value of the renormalization scale for the strong
coupling. We suggest to use this principle in a broader sense, requiring 
in our case the minimal sensitivity of the predictions to the change 
of both the renormalization and the energy scales, $\mu_R$ and $s_0$.

We replace for convenience in~(\ref{series}) $\ln(s/s_0)$ with $Y-Y_0$,
where $Y=\ln(s/Q^2)$ and $Y_0=\ln(s_0/Q^2)$, and study the dependence of the
amplitude on $Y_0$.

\begin{figure}[tb]
\centering
\hspace{-1cm}
{\epsfysize 8cm \epsffile{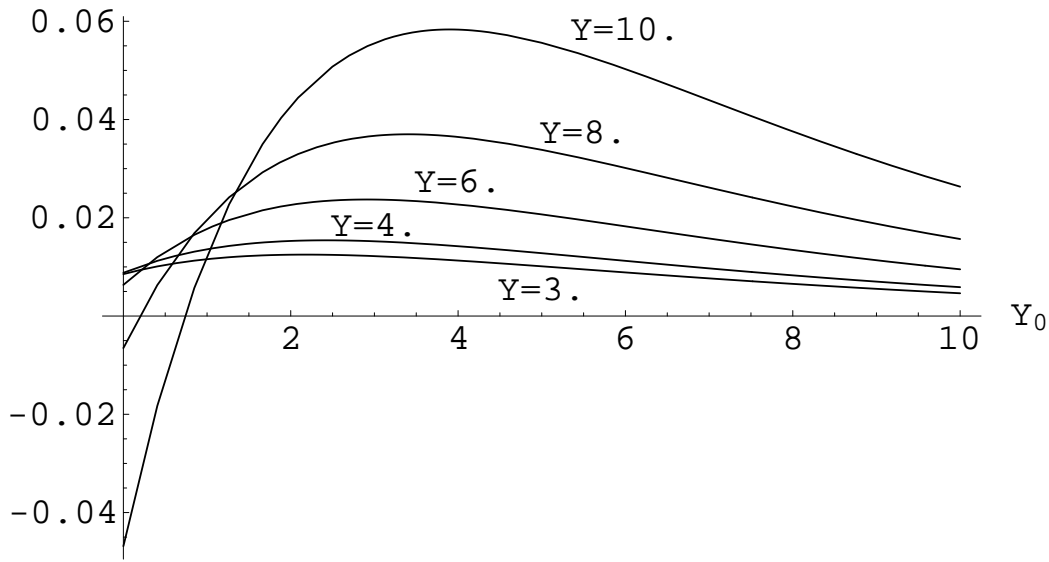}}
\caption[]{${\cal I}m_s ({\cal A})Q^2/(s \, D_1 D_2)$ as a function of
$Y_0$ at $\mu_R=10 Q$. The different curves are for $Y$ values of 10, 8, 6, 4 and 3.
The photon virtuality $Q^2$ has been fixed to 24 GeV$^2$ ($n_f=5$).}
\label{Y0}
\end{figure}

The next two figures illustrate the dependence on these parameters for
$Q^2$=24 GeV$^2$ and $n_f=5$.
In Fig.~\ref{Y0} we show the dependence of amplitude on $Y_0$ for
$\mu_R=10 Q$, when $Y$ takes the values 10, 8, 6, 4, 3.

\begin{figure}[tb]
\centering
\hspace{-1cm}
{\epsfysize 8cm \epsffile{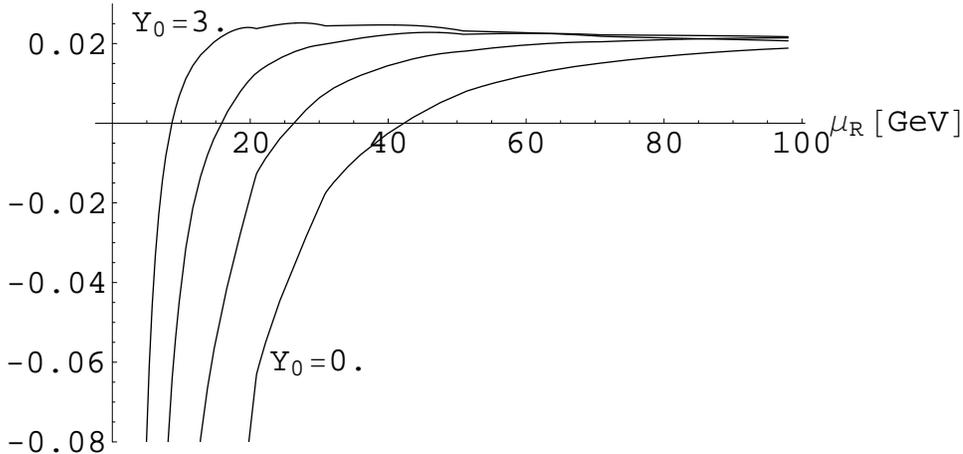}}
\caption[]{${\cal I}m_s ({\cal A})Q^2/(s \, D_1 D_2)$ as a function of
$\mu_R$ at $Y$=6. The different curves are, from above to below, for $Y_0$
values of 3, 2, 1 and 0. The photon virtuality $Q^2$ has been fixed to
24 GeV$^2$ ($n_f=5$).}
\label{muR}
\end{figure}

We see that for each $Y$ the amplitude has an extremum in $Y_0$ near which
it is not sensitive to the variation of $Y_0$, or $s_0$. Our choice of $\mu_R$
for this figure is motivated by the study of $\mu_R$ dependence. In
Fig.~\ref{muR} we present the $\mu_R$ dependence for $Y=6$; the curves from above
to below are for $Y_0$=3, 2, 1, 0.

\begin{figure}[tb]
\centering
\hspace{-1cm}
{\epsfysize 8cm \epsffile{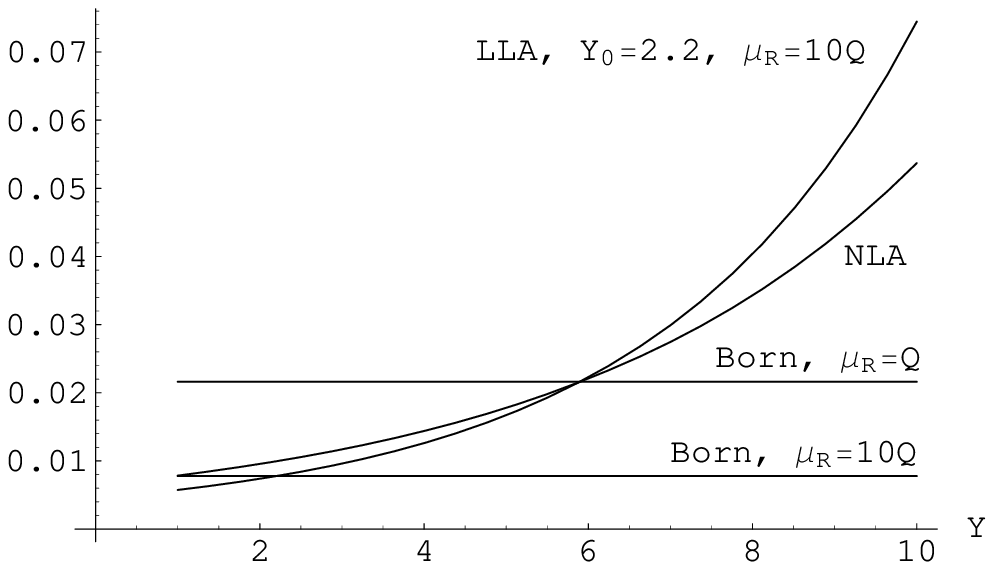}}
\caption[]{${\cal I}m_s ({\cal A})Q^2/(s \, D_1 D_2)$ as a function of
$Y$ for optimal choice of the energy parameters $Y_0$ and $\mu_R$ (curve 
labeled
by ``NLA''). The other curves represent the LLA result for $Y_0=2.2$ and $\mu_R=10Q$
and the Born (2-gluon exchange) limit for $\mu_R=Q$ and $\mu_R=10Q$.
The photon virtuality $Q^2$ has been fixed to 24 GeV$^2$ ($n_f=5$).}
\label{PMSres}
\end{figure}

Varying $\mu_R$ and $Y_0$ we found for each $Y$ quite large regions in
$\mu_R$ and $Y_0$ where the amplitude is practically independent on $\mu_R$ and
$Y_0$. We use this value as the NLA result for the amplitude at given $Y$.
In Fig.~\ref{PMSres} we present the amplitude found in this way as a function of
$Y$. The resulting curve is compared with the curve obtained from the LLA
prediction when the scales are chosen as $\mu_R=10 Q$ and $Y_0=2.2$, in
order to make the LLA curve the closest possible (of course it is not an exact
statement) to the NLA one in the given interval of $Y$. The two
horizontal lines in Fig.~\ref{PMSres} are the Born (2-gluon exchange)
predictions calculated for $\mu_R=Q$ and $\mu_R=10 Q$.

We stress that one should take with care BFKL predictions for small values
of $Y$, since in this region the contributions suppressed by powers of
the energy should be taken into account. At the lowest order in $\alpha_s$ such
contributions are given by diagrams with quark exchange in the $t$-channel and are
proportional in our case to $\alpha_{EM}\alpha_sf_V^2/Q^2$.
At higher orders power suppressed contributions contain double
logarithms, terms $\sim \alpha_s^n\ln^{2n} s$, which can lead to a significant
enhancement. Such contributions were
recently studied for the total cross section of $\gamma^*\gamma^*$
interactions~\cite{Bartels}.

If the NLA (and LLA) curves in Fig.~\ref{PMSres} are compared
with the Born (2-gluon exchange) results, one can conclude that the summation of
BFKL series gives negative contribution to the Born result for $Y<6$ if one chooses
for the scale of the strong coupling in the Born amplitude the value given by the
kinematics, $\mu_R=Q$. We believe that our calculations show that one should at
least accept with some caution the results obtained in the Born approximation,
since they do not give necessarily an estimate of 
the observable from below.

Another important lesson from our calculation is the very large scale for
$\alpha_s$ (and therefore the small $\alpha_s$ itself) we obtain using PMS.
It appears to be much bigger than the kinematical scale and looks unnatural
since there is no other scale for transverse momenta in the problem at
question except $Q$. Moreover one can guess that at higher orders
the typical transverse momenta are even smaller than $Q$ since they "are shared"
in the many-loop integrals and the strong coupling grows in the infrared.
In our opinion  the  large values of $\mu_R$ we found is not an indication of the
appearance of a new scale, but is rather a manifestation of the nature of the
BFKL series. The fact is that NLA corrections are large and then, necessarily,
since the exact amplitude should be renorm- and energy scale invariant, the
NNLA terms should be large and of the opposite sign with respect to the NLA.
We guess that if the NNLA corrections were known and we would apply PMS to the
amplitude constructed as LLA + NLA-corrections + NNLA-corrections,
we would obtain in such calculation more natural values of $\mu_R$.

In the last years strong efforts have been devoted to the
improvement of the NLA BFKL kernel as a consequence of the analysis of
collinear singularities of the NLA corrections and by the account of
further collinear terms beyond NLA~\cite{agustin}. 
This strategy has something in common with ours, in the sense that it is
also inspired by renormalization-group invariance and it also leads to
the addition of terms beyond the NLA. These extra-terms are large and of
opposite sign with respect to the NLA contribution, so that they partially
compensate the NLA corrections. The findings of the present work
suggest, however, that the corrections to the impact factors heavily 
contribute to the NLA amplitude, being even dominating in some interval
of non-asymptotically high energies. 
Moreover, by inspection of the structure of the amplitude in the regime of 
strongly asymmetric photon virtualities, one can deduce that also the
impact factors generate collinear terms which add up to those arising
from the kernel, see e.g. Eqs.~(84) and~(85) of Ref.~\cite{IKP04}. 
This leads us to the conclusion that in the approaches based on kernel 
improvement the additional information coming from impact factors should somehow be
taken into account when available. These issues 
certainly deserve further investigation and we believe that useful hints in this 
direction can be gained from the study of the $\gamma^* \gamma^* \to V V$ amplitude 
in the regime of strongly ordered photon virtualities.

We conclude this Section with a comment on the possible implications of
our results for mesons electroproduction to the phenomenologically 
more important case of the $\gamma^* \gamma^*$ total cross section. 
By numerical inspection we have found that the ratios $b_n/b_0$ we got for the 
meson case agree for $n= 1\div 10$ at $1\div 2\%$ accuracy level with the analogous 
ratios for the longitudinal photon case and at $3.5\div 30\%$ accuracy level 
with those for the transverse photon case. Should this similar behavior persist 
also in the NLA, our predictions could be easily translated to estimates of the
$\gamma^* \gamma^*$ total cross section.



\begin{thebibliography}{99}

\bibitem{IP05} 
D.~Yu. Ivanov and A. Papa, hep-ph/0508162.

\bibitem{BFKL}
V.S.~Fadin, E.A.~Kuraev, L.N.~Lipatov, Phys. Lett. {\bf B60} (1975) 50;
E.A.~Kuraev, L.N.~Lipatov and V.S.~Fadin, Zh. Eksp. Teor. Fiz. {\bf 71} (1976)
840 [Sov. Phys. JETP {\bf 44} (1976) 443]; {\bf 72} (1977) 377 [{\bf 45} (1977) 199];
Ya.Ya.~Balitskii and L.N.~Lipatov, Sov. J. Nucl. Phys. {\bf 28} (1978) 822.

\bibitem{gammaIF}
J.~Bartels, S.~Gieseke and C.~F.~Qiao, Phys. Rev. {\bf D63} (2001) 056014
[{\it Erratum-ibid.} {\bf D65} (2002) 079902];
J.~Bartels, S.~Gieseke and A.~Kyrieleis, Phys. Rev. {\bf D65} (2002) 014006;
J.~Bartels, D.~Colferai, S.~Gieseke and A.~Kyrieleis,
Phys. Rev. {\bf D66} (2002) 094017;
J.~Bartels, Nucl. Phys. (Proc. Suppl.) (2003) 116;
J.~Bartels and A.~Kyrieleis, Phys. Rev. {\bf D70} (2004) 114003;
V.S.~Fadin, D.Yu.~Ivanov and M.I.~Kotsky, Phys. Atom. Nucl. {\bf 65} (2002) 1513
[Yad. Fiz. {\bf 65} (2002) 1551];
V.S.~Fadin, D.Yu.~Ivanov and M.I.~Kotsky, Nucl. Phys. {\bf B658} (2003) 156.

\bibitem{NLA-kernel}
V.S.~Fadin and L.N.~Lipatov, Phys. Lett. {\bf B429} (1998) 127;
M.~Ciafaloni and G.~Camici, Phys. Lett. {\bf B430} (1998) 349.

\bibitem{IKP04}  
D.~Yu. Ivanov, M.I.~Kotsky and A. Papa, Eur. Phys. J. {\bf C38} (2004) 195; 
Nucl. Phys. (Proc. Suppl.) {\bf 146} (2005) 117.

\bibitem{PSW} 
B.~Pire, L.~Szymanowski and S.~Wallon, hep-ph/0410108, hep-ph/0501155, 
hep-ph/0507038.

\bibitem{PSW1}
R. Enberg, B.~Pire, L.~Szymanowski and S.~Wallon, hep-ph/0508134.

\bibitem{BRAUN}
P.~Ball, V.M.~Braun, Y.~Koike and K.~Tanaka, Nucl. Phys. {\bf B529} (1998) 323.

\bibitem{Stevenson}
P.M.~Stevenson, Phys. Lett. {\bf B100} (1981) 61; Phys. Rev. D {\bf D23} (1981) 2916.

\bibitem{Bartels}
J.~Bartels and M. Lublinsky, Mod. Phys. Lett. {\bf A19} (2004) 19691982.

\bibitem{agustin}
A.~Sabio-Vera, hep-ph/0505128 and references therein.

\end{thebibliography}
\end{document}